# Bandwidth enhancement and optical performances of multiple quantum well transistor lasers

Iman Taghavi[1, 2], Hassan Kaatuzian[1] and Jean-Pierre Leburton[2,3]

[1]*Photonics Research Laboratory, Electrical Engineering Department, Amirkabir University of Technology, Tehran 15914*

[2]*Beckman Institute for Advanced Science and Technology, University of Illinois at Urbana-Champaign, Urbana, IL 61801*

[3]*Department of Electrical and Computer Engineering, University of Illinois at Urbana-Champaign, Urbana, IL 61801*

A detailed rate-equation-based model is developed to study carrier transport effects on optical and electrical characteristics of the Multiple Quantum Well Heterojunction Bipolar Transistor Laser in time-domain. Simulation results extracted using numerical techniques in small-signal regime predict significant enhancement in device optical bandwidth when multiple quantum wells are used. Cavity length and base width are also modified to optimize the optoelectronic performances of the device. An optical bandwidth of ≈60GHz is achieved in the case of 5 quantum wells each of 70Å widths and a cavity length of 200μm.

The observation of stimulated light emission (λ≈900nm) from the quantum well (QW) base region of the high-speed heterojunction bipolar transistor laser[1] (HBTL) has received significant attention in recent years. With a highly doped p-type base region, the transistor laser (TL) is expected to have a short radiative recombination lifetime that would make it faster than conventional optical sources for communication networks, i.e. diode lasers (DL). Thanks to its double functionality, i.e. electrical switching of the HBT and lasing operation, the TL has significant potential for ultrafast optoelectronic integrated circuits. As it is anticipated that the 3-terminal optoelectronic device will have a better optical bandwidth compared with a DL, optimization of QW-base structure is desirable.[2] Experimentally the optimization of the TL structure without taking into account the HBT response would be tedious, whereas only a few theoretical models have been developed for single quantum well (SQW) HBTL, for which simplified forms of conventional rate equations for DL have been implemented.[3,4,5]

Recently a large signal analysis has been proposed[6] that simulates optical and electrical performances of a buried heterostucture TL with different number of QWs. However, this approach suffers from several drawbacks such as linear optical gain, which is only a good approximation for small signal analysis, and the neglect of tunneling effects between different QWs. In their work, the authors reported a recombination lifetime in the TL active region of the order of a few nanoseconds, which is comparable to DL lifetime. In this work, by using a comprehensive rate equation model we predict dependence of the TL performances on the structural factors of the waveguide region, e.g. base and collector width, number, width, location and material of QWs and material systems. Specifically, we show that optical performances can be enhanced with optimization of the QW base of the HBTL. For this purpose, we calculate device parameters as a function of above-mentioned factors. We focus on DL integration with a HBT, in which the high speed electrical switching of DL is combined with fast stimulated recombination in the HBT active region. We include separate confinement heterostructure (SCH) regions, tunneling, thermionic escape and capture lifetimes to study the carrier transport effects, appropriately. Furthermore, in order to assess the carrier dependent gain in a large signal analysis, bulk base and quantum well active region recombination lifetimes are calculated as a function of both minority and majority carrier concentrations. An effective optimization procedure can be performed on the different effects of the above-mentioned structural factors on various optical and electrical characteristics of the TL, e.g. optical bandwidth and output power, beam quality, current gain, threshold current, etc.

Figure 1 illustrates the carrier transport process in a multiple quantum well (MQW) HBTL. The base region consists of three sub-regions: a) two SCH regions that are not necessarily equal in width, i.e. unlike in DL, b) the QWs and c) the barriers. The carriers are assumed to be captured by the first QW, and then transported to the next well by tunneling (for barriers thinner than 135Å[7]) or thermionic emission and diffusion across the barrier. For this purpose, we develop a rate-equation-based model for an N-p$^+$-i HBTL with an arbitrary number of QW. In order to account for slow inter-well carrier transport in MQW structure, we assigned different values to the density of carriers in each well. Additionally, we differentiate between QW and bulk base carriers. In our approach, we normalize the carrier density of each SCH region as well as each QW according to their relational volume, whereas the photon density is normalized with respect to the volume of all the wells. It is assumed that the carriers diffuse from emitter to the left SCH region of the base then travel to QWs. The coupled rate equations for MQW-HBTL are

$$\frac{dN_{sch}}{dt} = \frac{\eta_i J}{qW_{sch1}} - N_{sch}\sum_{z=1}^{M}\frac{\upsilon_z}{\tau_{cap,z}} - \frac{N_{sch}(1-\sum_{z=1}^{M}\upsilon_z)}{\tau_{nsch}} - \frac{N_M}{\tau_e}\frac{W_{qw}}{W_{sch2}} \quad (1)$$

$$\frac{dN_1}{dt} = N_{sch}\frac{\upsilon_1}{\tau_{cap,1}}\frac{W_{sch1}}{W_{qw}} - \frac{N_1}{\tau_e} - \frac{N_1}{\tau_{nqw}} - \frac{N_1 - N_2}{\tau_c} - \upsilon_g g(N_z)(1-\varepsilon\Gamma S)S \quad (2)$$

$$\frac{dN_z}{dt} = N_{sch}\frac{\upsilon_z}{\tau_{cap,z}}\frac{W_{sch1}}{W_{qw}} + \frac{N_z - N_{z-1}}{\tau_c} - \frac{N_z}{\tau_e} - \frac{N_z}{\tau_{nqw}} - \frac{N_z - N_{z+1}}{\tau_c} - \upsilon_g g(N_z)(1-\varepsilon\Gamma S)S \quad (3)$$

$$\frac{dN_M}{dt} = N_{sch}\frac{\upsilon_M}{\tau_{cap,M}}\frac{W_{sch1}}{W_{qw}} + \frac{N_M - N_{M-1}}{\tau_c} - \frac{N_M}{\tau_e} - \frac{N_M}{\tau_{nqw}} - \upsilon_g g(N_M)(1-\varepsilon\Gamma S)S \quad (4)$$

$$\frac{dS}{dt} = \Gamma\upsilon_g\left[\sum_{z=1}^{M}g(N_M)\right](1-\varepsilon\Gamma S)S - \frac{S}{\tau_p} + \frac{\beta}{M}\sum_{z=1}^{M}\frac{N_z}{\tau_n(N_z)} \quad (5)$$

where $J$ is the injected current density, $q$ is the electron charge, $\eta_i$ is the internal quantum efficiency, $W_{sch1}$ and $W_{sch2}$ are width of left and write SCH regions, respectively, $W_{qw}$ is the well width, $M$ is the number of quantum wells, $N_{sch}$, $N_1$, $N_z$ and $N_M$ are the carrier density in total SCH region, first, inner and the last QWs correspondingly, $\Gamma$ is the optical confinement factor of the total waveguide region, $g(N_z)$ is the carrier-density dependent gain of the $z$th QW, $\varepsilon$ is the gain compression factor, $S$ is the photon density normalized with respect to QW volume, $\upsilon_g$ is the group velocity and $\beta$ is the spontaneous emission factor. The QW geometry factor for the $z$th well, $\upsilon_z$, is defined as $\upsilon_z = W_{qw}/W_B \times (1 - x_{qw,z}/W_B)$, which is based on the carrier capture concept introduced in a previous model.[8] The first and second terms in the right hand side of the equation account for effects of relational thickness and location of the corresponding QWs, respectively. The location of $z$th well, $x_{qw,z}$, is calculated as $x_{qw,z} = W_{sch1} + (z-1)\times(W_{qw}+l_b)$ where $l_b$ is the barrier width.

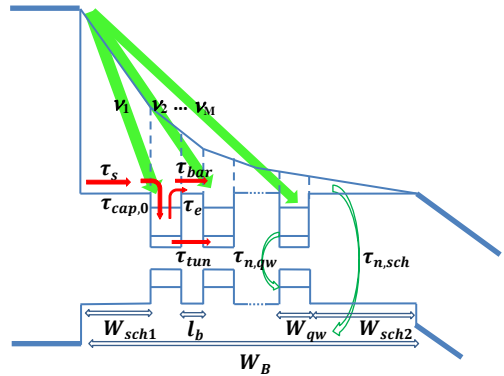

FIG. 1. (Color online) Schematics of carrier transport model for MQW-HBTL.

In our model, we use a nonlinear, logarithmic carrier dependent gain in order to make the model appropriate for both small and large signal analysis[9]

$$g(N_z) = G_0 \ln(\frac{AN_z + BN_z^2 + CN_z^3}{AN_0 + BN_0^2 + CN_0^3}) \quad (6)$$

where $N_z$ is the density of carriers in the correspondence QW, $N_0$ is transparency carrier density and $G_0$ is the gain coefficient. A tenth-order Taylor series is utilized for gain in numerical computation. The gain parameters are extracted from common equations for material gain.[10,11]

A crucial difference between our rate equation model and DL models[12] is the presence of the direct carrier capture in the **z**th QW, z=2 to M, described by the first term in the right hand side of the equations 3-4. As a result, due to the carrier flow towards the reversely bias collector-base junction, there is a larger current of carriers captured into the successive QWs, z=2 to M, of the MQW-HBTL than in DLs. Moreover, carrier sweeping through the base, which prevents charge build-up, enhances the differential capture rate per QW in the MQW-HBTL. Independently of slow inter-well carrier transport, the enhancement due to this effect is more significant for the HBTL with increasing number of QWs, and results in a wider optical bandwidth for the TL than for the DL.

The overall capture time for the first QW, $\tau_{cap,1}$, is defined as the sum of carrier transport over the left SCH region, $\tau_s$, and the finite capture time to virtual states, $\tau_{cap,0}$. Carrier transport across the SCH region is related to SCH width and diffusion constant ($D$) by $W_{sch}^2/2D$. The diffusion constant is independently calculated by the method of Klausmeier-Brown et al.[13] For other QWs, i.e. $\tau_{cap,z}$ where z=2 to $M$, it is also necessary to account for transport time over barriers. $\tau_e$, $\tau_p$ are thermionic emission and photon lifetimes, while $\tau_c$ is defined as effective inter-well transport time that we calculated for TL as $\tau_c = 1/[1/\tau_{tun} + 1/(\tau_e + \tau_{bar} + \tau_{cap,0})]$. The $\tau_{nsch}$ and $\tau_{nqw}$ are carrier density dependent recombination lifetimes in SCH and QW regions, respectively that are calculated using standard $A$, $B$ and $C$ recombination coefficients, i.e. SRH, radiative and Auger recombination processes (data not shown).[14,15] The internal quantum efficiency incorporated in the model is computed for the device at threshold current.[16]

Given the special design of the TL optical waveguide compared to that of DL, we modified Γ for each number of QW[7] as formulated by Botz[17] and Streifer et al.[18] Like other device parameters, Γ depends on structural factors and may not be estimated by a simple relationship amongst the number of the QWs due to the asymmetric waveguide design of the HBTL. Although the model is general enough to be applicable to MQW-HBTL with different materials, in this analysis we consider the $Al_{0.4}Ga_{0.6}As/GaAs/GaAs$ system for the three HBT emitter ($A≈1000μm^2$), base and collector regions, respectively. The base region consists of strained $In_{0.25}Ga_{0.75}As$-GaAs QWs each of 70Å width sandwiched between GaAs barriers with 80Å thicknesses. The active region is located in the middle of the base region. TL is biased in common base configuration assuming $I_B$=60mA, $I_C$=375mA for SQW, $L$=200μm and facet reflectivity of $R≈0.32$, throughout this paper unless otherwise stated.

Figure 2 shows our simulation results for Γ as a function of barrier and base widths (as part of total optical waveguide region). As can be seen, for each barrier width Γ exhibits a maximum for an optimum base width. In addition, Γ is directly proportional to the barrier width because of its direct impact on the effective refractive index of the optical waveguide. The inset in figure 2 shows a comparison between optical confinement factor of the HBTL with a SQW and 3QW structures. Optimum base width consequently depends on the number of QWs. Assuming that in the TL active layer loss is dominant, the total internal loss is calculated as $\Gamma \alpha_{active} + (1-\Gamma)\alpha_{clad}$ where $\alpha_{active}$ and $\alpha_{clad}$ are active and cladding layers loss respectively.

In our model, the DC and AC small signal performances of the device are calculated using the floating-point arithmetic, while either the Rosenbrock or Fehlberg fourth-fifth order Runge-Kutta numerical methods are utilized to simulate large signal analysis. These computationally efficient numerical methods are utilized together with pre-defined initial conditions to avoid solution divergence. A recursive method is also necessary for calculation of the internal quantum efficiency. Due to very small values used in the Ordinary Differential Equation (ODE) system, it is more efficient to utilize non-zero initial carrier concentrations in our computational approach so that the solutions converge rapidly. An optimization of MQW-HBTL optoelectronic characteristics, however, can be performed using the implicit Rosenbrock third-fourth order Runge-Kutta method that is computationally more stable for stiff ODE system.[19] The carrier transport model is solved for DC and subsequently AC analysis. Thermal effects on the device parameters and performances will be considered in a forthcoming paper.

Figure 3 displays simulation results for intrinsic optical modulation response of HBTL's with different QW numbers. The bandwidth increases for the MQW-HBTL when the number of QWs rises until it reaches a maximum of ≈60GHz for five QWs and then decreases for more wells. However, this optimum number of QW depends on other structural factors as well as biasing condition. Maximum overshoot of the response increases constantly up to ≈6dB for the case of six wells. In addition, a comparison between modulation response of the DL and TL with SQW and 5QW structures is shown in inset. Another inset of figure 3 displays simulation results for the MQW-HBTL optical bandwidth compared to that of a MQW-DL as a function of QW number. As previously predicted, bandwidth enhancement is more pronounced in TL.

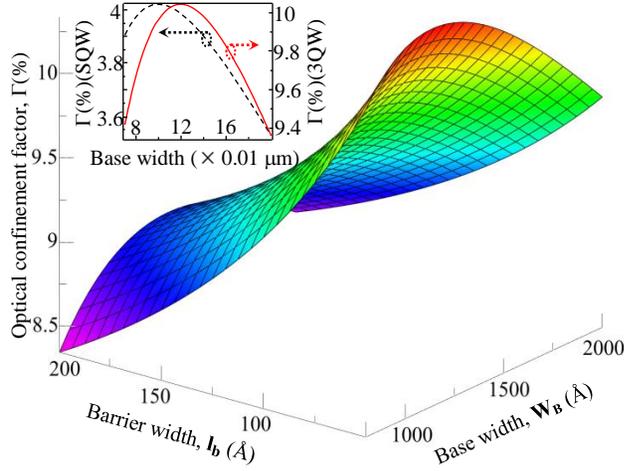

FIG. 2. (Color online) Simulated optical confinement factor of 3QW-HBTL. The inset compares Γ between SQW and 3QW structures calculated individually. Maximum value for Γ is obtained by selecting appropriate base width and barrier width for each number of QW.

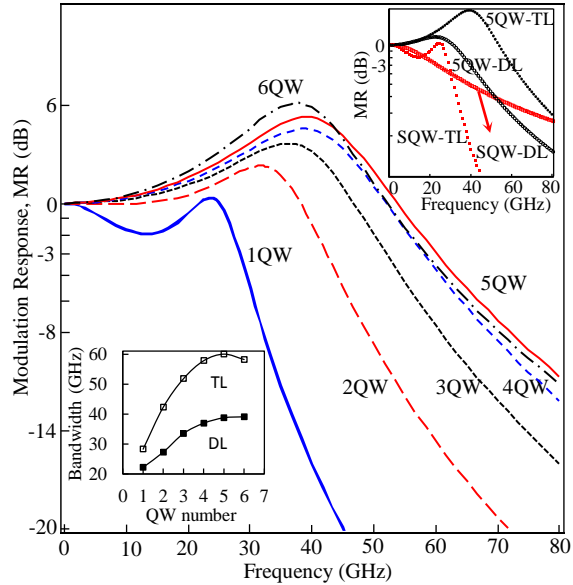

FIG. 3. (Color online) Simulated intrinsic optical modulation frequency response for HBTL with different number of QWs. An optimum bandwidth of ≈60GHz with overshot peak of 5dB is achievable for 5QW. The upper (lower) inset compares modulation response (optical bandwidth) between TL and DL using SQW and 5QW structures (versus QW number).

Bias dependent optical performances of SQW and MQW structures up to six QWs are shown in figure 4 in which the optical bandwidth varies versus the base current. Under low current level, the SQW structure has a superior bandwidth to the MQW device due to the higher rate of carriers that are captured by the single QW. The bandwidth increases up to a saturation value of ≈29GHz because of the saturation in material gain. In the MQW case, bandwidth enhancement starts in relatively higher injection levels for which each of the wells is filled with enough carriers.

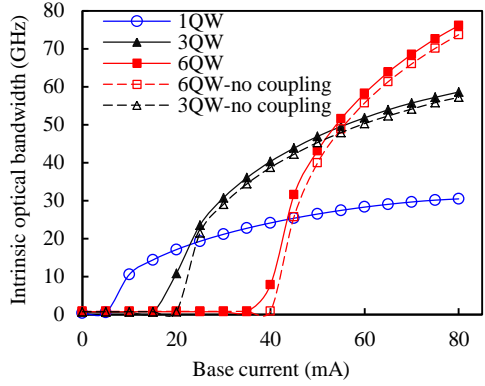

FIG. 4. (Color online) Calculated optical modulation bandwidth for HBTL with up to 6 wells versus base current. MQW structure has a better optical performance than the SQW in higher bias currents. For comparison purpose, we also showed simulated bandwidth of HBTL with 3 and 6 QWs without coupling effects.

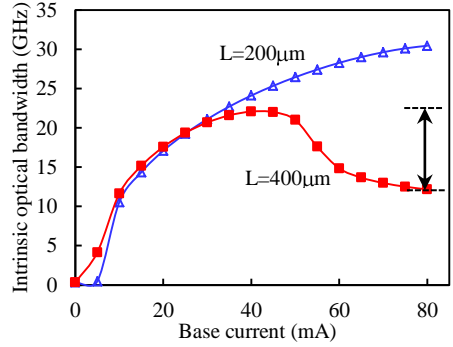

FIG. 5. (Color online) Bandwidth saturation and drop-off effects due to carrier transport in SQW-HBTL. Shorter cavity length is needed to avoid abrupt decrease in bandwidth under high injection level.

Bandwidth saturation and drop-off is more evident in figure 5, where the optical bandwidth is sketched for a SQW-HBTL for two different cavity lengths. This effect, which is most likely caused by carrier transport effects described before, sets the limit for the device performances. This is in good agreement with the experimental data for the SQW and MQW diode laser reported by Nagarajan *et al*.[20] However, the difference arises from the fact that they have calculated the bandwidth as a function of output power rather than the current.

Calculation of the differential current gain ($\beta = \Delta I_C / \Delta I_B$) in conjunction with our previous results on optical bandwidth illustrates the gain-bandwidth trade-off in the MQW-HBTL predicted before.[21,22] Figure 6 shows the mentioned trade-off between carrier collection in optical and electrical collectors where $\beta$ is considerably lower for MQW compared with SQW structure as a result of higher recombination rate in multiple wells. We simulated the inter-well coupling effects on optical performances of the MQW-HBTL as shown in figure 8. Apparently, this effect can increase the optical bandwidth by a few GHz because of fast redistribution of carriers between subsequent wells as well as slightly different carrier-dependent gain profile. The influence of other important structural factors on the HBTL optoelectronic characteristics can be obtained by our carrier transport model, as seen in figure 7 that displays the modulation bandwidth dependence on the cavity length ($L$) for different structures. For each number of QWs, there is an optimum $L$ for which the bandwidth is maximized. The optimum cavity length is 150μm for the SQW device compared with 150μm for 5QW-HBTL, which is difficult to implement. The trade-off between carrier dependent differential gain and photon lifetime is responsible for this maximum cavity length. In the other hand, bandwidth sensitivity to cavity length is higher for SQWs. This is somewhat in contrast to the situation in DL. We believe that the discrepancy arises from the previously mentioned dual functionality of the transistor laser. The locus of optimum L is also indicated in this figure.

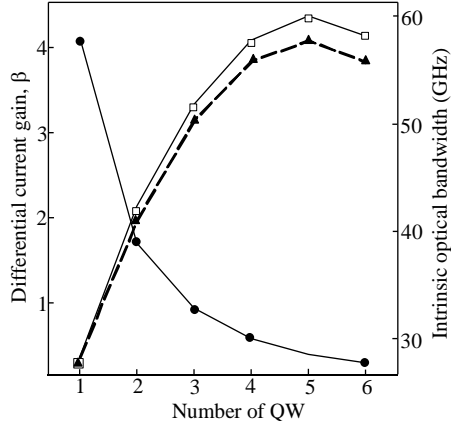

FIG. 6. Optical bandwidth-Current gain trade-off for HBTL versus number of QWs. SQW structure has a higher β while 5QW-HBTL has the maximum bandwidth. The optical bandwidth with (solid line) and without (dash line) effects of inter-well coupling have been shown on the figure.

In conclusion, our model for device structure optimization anticipates that the HBTL optical bandwidth can be enhanced up to 60GHz when 5QWs, each 70Å wide are incorporated within the base active region inside a 200μm cavity length. The originality of our work relies on the fact that we independently calculate optical and electrical parameters of HBTL for different structures and arbitrary number of wells. Our detailed carrier transport model for MQW-HBTL has the capability to simulate the gain-bandwidth trade-off behavior of the device as well as drop–off and saturation of optical bandwidth in comparison with SQW structure. We also obtained the bias dependent optical characteristic of the MQW-HBTL.

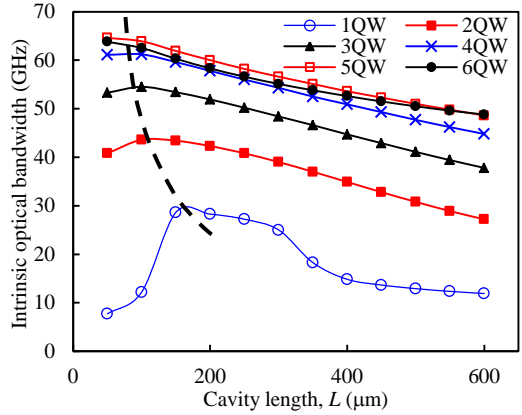

FIG. 7. (Color online) HBTLs with MQW have larger modulation bandwidth and less sensitivity than SQW but shorter cavity length is needed to take full advantage of this effect.

The authors would like to thank the Beckman Institute and University of Illinois for support. One of the authors (I. T.) is also grateful for the support of the Ministry of Science, Research and Technology of Iran.